\documentclass[10pt,a4paper]{article}
\usepackage[utf8]{inputenc}
\usepackage[T1]{fontenc}
\usepackage{amsmath,natbib}
\usepackage{amssymb}
\usepackage{makeidx,subcaption}
\usepackage{graphicx,color}
\usepackage{comment}

\begin{document}
\title{Bifractal behavior of Solar supergranulation and magnetic activity}
\author{ Rajani G$^1$, Sowmya G M$^2$, U Paniveni$^{3,4}$, R Srikanth$^4$}
\date{}
\maketitle
$^1$ PES College of Engineering, Mandya - 571401, Karnataka, India.\\
$^2$ GSSS Institute of Engineering and Technology for Women, KRS Road,Metagalli Mysuru-570016, Karnataka, India\\
$^3$ Bangalore University, Jnanabharathi, Bengaluru – 560056\\
$^4$ Poornaprajna Institute of Scientific Research, Devanahalli, Bangalore-562110, Karnataka, India\\
\begin{abstract}
We study the complexity and scale of the supergranular network across the 23rd solar cycle, using the Ca II K digitized intensitygrams from the Kodaikanal Solar Observatory (KSO). Enhancing our previous data and refining our data analysis, we study supergranular fractal dimension as a function of cell size. We find that across the cycle phases, the cells show a bifractal behavior, with approximately half the larger cells in the studied scale range showing a slightly greater fractal dimension than the smaller cells. We also study the discrepancy between supergranular scale as determined by direct inspection methods (around 17 Mm) and autocorrelation (around 30 Mm), and attribute this to a preferential selection of well defined cells in the former case.
\end{abstract}

\section{Introduction}
Supergranulation, believed to be a convective phenomenon, is a solar surface network of cellular horizontal velocity pattern with a typical size of 30 Mm and a lifetime of one day. Supergranulation is believed to be feature of the convection in Sun-like stars.  Typically, it is most pronounced in the Ca II K 3934 \AA ~line \citep{chatzistergos2022full}, but can also be  deduced from magnetograms or Dopplergrams (Simon and Leighton 1964; Singh and Bappu 1981; Hagenaar, Schrijver et al. 1997; Gontikakis, Peter and Dara).

Supergranular scales may be extracted by direct inspection \citep{sowmya2022supergranular, rajani2022solar}, or by automated techniques such as autocorrelation \citep{simon1964velocity, srikanth2000distribution, badalyan2008quasi}, which is related to the width of the network boundary \citep{patsourakos1999transition}. The magnetic fields hemmed in by supergranular horizontal motion to the boundaries are understood to rise above and form a magnetic canopy that expands right into the coronal layer \citep{reeves1976euv, gabriel1976magnetic, patsourakos1999transition}. 

Interestingly, the length scales of the transition region as observed using the Coronal Diagnostic Spectrometer (CDS) instrument of Solar and Heliospheric Observatory (SOHO) show large asymmetry \citep{harrison1995coronal}. The magnetic fields of the Solar atmopshere may lead to turbulence, as indicated by Hinode experiment \citep{centeno2007emergence, lites2008horizontal, suarez2008magnetic, schussler2008strong}, suggestive of a small-scale turbulent dynamo in the photosphere \citep{boldyrev2004magnetic, vogler2007465, graham2009turbulent}. 

Fractal dimension $D$ provides a mathematical tool to quantify the complexity of geometric patterns, via a measure of the 
self-similarity in the pattern \citep{hutchinson1981fractals}. In the context of Solar supergranulation, $D$ provides a window on the magneto-convective turbulence in the Solar plasma, a point we elaborate on in the final Section. Given a plot of supergranular area $A$ and perimeter $P$, the dimension can be read off according to
\begin{equation}
A \propto P^{2/D}
\label{eq:FD}
\end{equation}
A regular cellular pattern corresponds to $D = 1$, while a frazzled, irregular one that is space-filling corresponds to $D > 1$.  In the context of Solar surface phenomena, \citet{muller1987dynamics} reported a fractal dimension of Solar granulation of $D = 1.25$ and $D\approx2$ for smaller and larger granules, respectively. \citet{paniveni2005fractal} reported a supergranular fractal
dimension of $D \approx 1.24$ for the Solar and Heliospheric Observatory (SoHO) dopplergrams, which was the basis of subsequent analysis of the Solar magneto-convective turbulence \citep{paniveni2011solar}. 

According to the Kolmogorov hypothesis applied to convection in a turbulent plasma, the horizontal velocity of a convective cell scales as $v_{\rm h} \propto L^{\frac{1}{3}}$, where $v_{\rm h}$ is the horizontal speed and $L$ the size of the convective cell \citep{krishan2002relationship}.  A direct relationship between magnetic activity and cell fractal dimension has been reported \cite{nesme1996fractal,rincon2018sun}. On the other hand, the interplay with the Solar cycle phase appears to modify the picture. Specifically, \citet{chatterjee2017variation} report that whereas active-region supergranular fractal dimension is lower than that for quiet regions, the former cells may have lower fractal dimension during the active phase, and the latter supergranules contrariwise. 

These considerations suggest that the interplay of Solar cycle phase, magnetic activity and the supergranular plasma outflow contribute to determining the fractal dimension, scale or any other cell parameter. Here, one concerns the relation between supergranular size and fractal dimension over the course of the Solar cycle. This was partialy addressed in our previous work \citep{rajani2022solar}, employing the method of analysis used by \citet{paniveni2005fractal}. The above work by \citet{rajani2022solar} is limited to the quiescent region,  and qualitatively indicated that there seemed to be a variation of the fractal dimension as a function of scale. This issue is now taken up in quantitative detail in the present work, where we report a bifractal behavior of supergranulation.

The remaining article is structured as follows. In Section \ref{sec:data}, we introduce the data and analytical tools used. Our result on the bifractal nature of supergranulation is reported in Section \ref{sec:results}, where for brevity we only present the detailed data for active phase. Here we also the discuss the discrepancy on cell size as estimated by the autocorrelation and direct image inspection methods. Finally we discuss in Section \ref{sec:conclusions} a simply model based on Kolmogorov theory for the observed bifractal behavior, and also present our conclusions.

\section{ Data and Analysis \label{sec:data}}
Data from the 23rd solar cycle (August 1996 -- December 2008) at KSO was employed for the analysis here. The KSO  
houses a K-line spectroheliograph 
with a  spectral dispersion of 7 \AA mm$^{-1}$ near 3934 \AA. It uses a 6 cm image obtained out of a Cooke photovisual triplet of 30 cm . A 46-cm diameter  Foucault siderostat reflects sunlight onto a lens of 30 cm. Exit slits centered at K 232 admit light at 0.5 A. The images are digitized into strips that run parallel to the Solar equator. The image resolution is 2$^{\prime\prime}$, twice the granular scale. The data is time-averaged over a 10-minute window in order to wash out any signatures of the 5-minute oscillations  \citep{paniveni2004relationship}. The contribution of granular velocity is lowered to a good extent by virtue of the spatial resolution and time averaging. Likewise, the signal due to p-mode vibrations is lowered through time averaging \citep{paniveni2005fractal}. 

The  activity level determines the extent of magnetic field dispersal, and thereby also affects the properties of the supergranular cell network, specifically its degree of irregularity, as quantified using a Mandelbrot fractal analysis. We expect such properties to continuously vary across the cycle. For our data, this behavior can be broadly captured by dividing the cycle into three phases-- peak, minimum and intermediate. In the considered cycle, we identify three phases: the peak phase (period 2000--2002), which features large sunspots and other signatures of high activity, such as Solar flare and coronal mass ejections; the minimum phases on either side (periods 1996--1997 and 2005--2007), during which such manifestations of magnetic activity are minimal, and the remaining period corresponding to the intermediate phase.

In the present work, we analyzed the multifractal nature of supergranulation during different epochs of the Solar cycle based on the above mentioned activity levels. We quantitatively find a scale-dependent multifractal behavior, by investigating an idea proposed earlier by \citet{rajani2022solar}, essentially by extending and refining the data analysis there . Specifically, there is a small difference between the fractal dimension for smaller and larger cells in a given phase, and furthermore, this trend varies with the phases of the solar cycle.  Examining the  cell's  perimeter-area  relationship  we derive the fractal dimension $D$ via the relationship of Eq. (\ref{eq:FD}), building on our previous work  \citep{sowmya2022supergranular}. 

Another related investigation concerns the  size  distribution  of supergranular cells, wherein we discuss the marked difference between cell sizes obtained by a tessellation technique or direct visual inspection of well defined cells, typically  16--18 Mm \citep{hagenaar1997distribution, rajani2022solar} vs automated methods such as a autocorrelation, typically around 30 Mm \citep{srikanth2000distribution}.

For the autocorrelation analysis, assuming that the network characteristics do not change significantly during a three-minute period, we have selected the central window to minimize the impact of foreshortening effects.. Also, the low-latitude location of our region of interest means that the horizontal-vertical asymmetry in supergranular length scale at higher latitudes \citep{raju2020asymmetry}  does not affect us. From selected images, rectangular windows of size (10 $\times$ 120) pixels were taken in the central meridian for which the two-dimensional autocorrelation function as a function of lag in the equatorial direction was calculated using Matlab. The length scale is estimated as twice the halfwidth of the autocorrelation function. To improve statistical robustness, the process was repeated for 20 images in each year of the cycle, and mean and the standard deviation of the length scales were obtained for error analysis purposes. 

\section{Results \label{sec:results}}
 
\begin{figure}
	\centering
	\includegraphics[width=.9\linewidth]{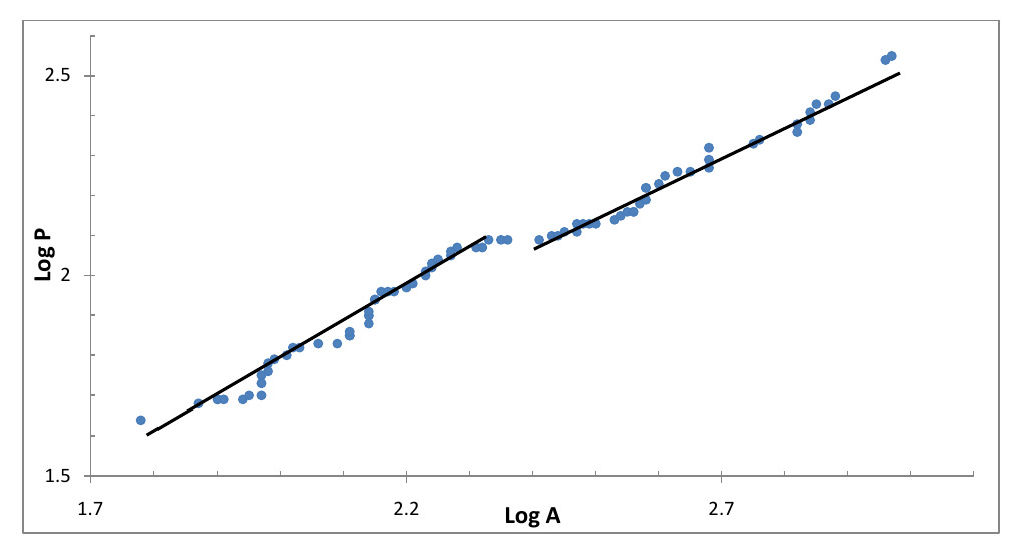}
	\caption{ Plot of $\log(A)$ v/s $\log(P)$ for supergranules in a quiet region of the Sun during the minimum phase of Solar cycle 23. The point of inflection where the fractal dimension increases slightly is around $\log A = 2.4$ Mm$^2$. For details pertaining to all three phases, refer to Table \ref{tab:inflection}.}
	\label{fig:inflection}
\end{figure}

Our result indicates a rather broad spread of length scales around a characteristic scale.  In each activity level, we find  the  existence  of a `critical' area at which a change in the geometrical properties, characterized by the fractal dimension, occurs. On the area vs perimeter plot, this corresponds to a point of inflection in the slope. The plot for the minimum phase is given in Figure \ref{fig:inflection}).  Similar plots are obtained for the other two cycle-phases. The results for all three phases is given in Table \ref{tab:inflection}.


\begin{table}
	\begin{tabular} {| p {1 cm} | p {2.5 cm} | p {2 cm} | p {2 cm} | p {2 cm} | }
		\hline
		\textbf{Sl.No.}&  	\textbf{phases} & 	\textbf{Point of inflection} & 	\textbf{Fractal dimension} & 	\textbf{weighted average goodness-of-fit}\\
		\hline
		1 & Minimum & A$\le$ 2.4  & 1.6 & 0.973\\ 
		\cline{3-4}
		&&A$\ge$ 2.4&1.68&\\
		\hline
		2 & Intermediate & A$\le$ 2.6 & 1.3 & 0.976\\
		\cline{3-4}
		&&A$\ge$ 2.6&1.6&\\
		\hline
		3 & Peak & A$\le$ 2.25 & 1.0 & 0.982\\
		\cline{3-4}
		&&A$\ge$ 2.25&1.2&\\
		\hline
	\end{tabular}
	\caption{Bifractal behavior of supergranulation based on the visual inspection analysis, showing a differentiation between smaller and larger supergranules in each activity phase. The inflection point shows a slight phase dependence, being largest for the intermediate phase.}
	\label{tab:inflection}
\end{table}

This point is determined carefully by sampling different divisions of the data set, to ensure that the selected two line segments correspond to maximal fits to the data points in the segment. The weighted average of the goodness of fit of the two segments is given in the last column of Table \ref{tab:inflection}. The obtained fractal dimension data supports a bifractal structure for all three phases. The point of inflection is determined by evaluating the point that maximizes the weighted goodness of fit. 

Calcium network diameters are shown to be smaller by 5$\%$ at solar maximum than at minimum \citep{singh1981dependence}.The average cell size is larger than the earlier reported value of 17--23 Mm derived by visual inspection, though sizes obtained by autocorrelation can indeed be larger \citep{singh1981dependence}.According to the turbulent convection theory. The distance between the barycentre of adjacent cells is described by a Gaussian distribution characterized by mean intercell distance of 24Mm and a FWHM of 5Mm \citep{berrilli1997time}. 

\begin{figure}
	\includegraphics[width=1\linewidth]{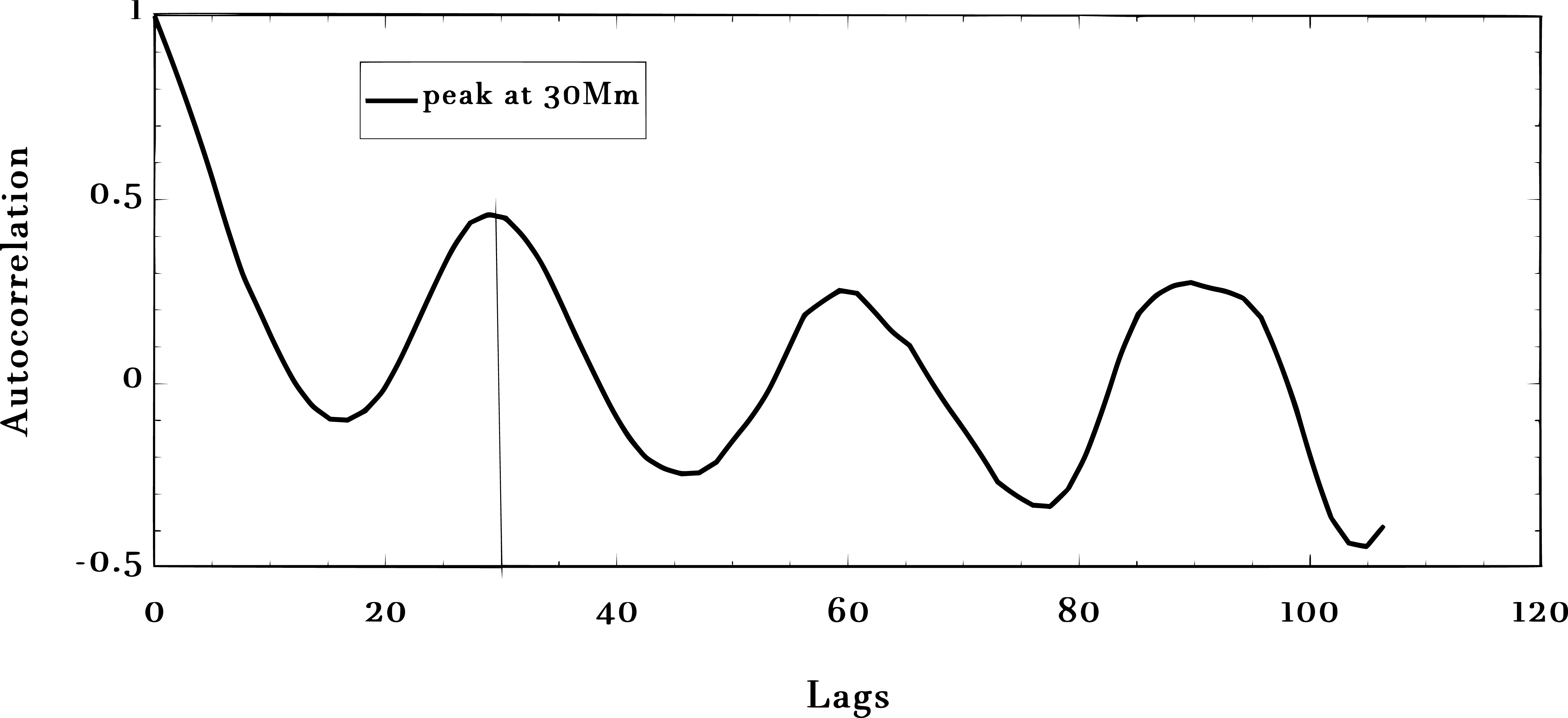}
	\caption{Plot of Autocorrelation function against lag which gives the length scale of 29 Mm.}
	\label{fig:autocorrelation1}
\end{figure}

In this work, we determined length-scales employing both the autocorrelation and visual inspection methods. A typical autocorrelation function plot for our data is depicted in Figure \ref{fig:autocorrelation1}. For the latter method, following the procedure described by \citet{rajani2022solar, sowmya2022supergranular}, we  pick up well defined small cells to delineate adjacent supergranular cells and analyze the intercellular distance between consecutive cells. By this method, we obtain scales that can well smaller than 20 Mm, as we reported in the above works. This is also consistent with the scales reported by \citet{hagenaar1997distribution}. Thus, for the same data, we find a discrepancy between the visual inspection and autocorrelation technique in finding the cell size.

\section{Discussions and conclusions \label{sec:conclusions}} 
The phase-dependence of supergranulation scales and fractal dimension were studied using the KSO Ca II K images of the 23rd cycle (years 1997 to 2007), building on earlier work \citep{rajani2022solar}. Both show a dependence on the solar cycle activity level. 

Scales were estimated using both autocorrelation and visual inspection. The relatively larger value of the former is attributed to a possible selection bias towards smaller cells, in the latter method. Since only those cells with well defined boundaries were chosen, and usually larger cells have boundaries that more irregular and sometimes broken, whereas the autocorrelation method is indifferent to such shape factors, this could explain the possible bias. Here we may note that the autocorrelation technique provides an average scale, rather than a realistic one \citep{srikanth1999chromospheric}, that too, possibly an overestimate \citep{singh1981dependence}. 

In an earlier work by two of the present authors on Solar dopplergrams \citep{krishan2002relationship}, it was reported that the the horizontal speed $v$ of the supergranular outflow depends on the length scale $L$ according to the observed function $v \sim L^{1/3}$. This is consistent with the application of Kolmogorov theory for a turbulent medium to supergranular convection. Accordingly, assuming that the plasma is a turbulent magneto-convective medium, the fractal dimension $D$ can serve as a tool to probe properties of the turbulent magneto-convection in the Solar plasma associated with supergranuation. Specifically,
 we can determine $D$ for different \textit{isosurfaces}, i.e., surfaces of constant temperature, concentration or pressure \citep{mandelbrot1975geometry}.  It can be shown as a consequence of the Kolmogorov energy spectrum, $k^{-5/3}$, that the variances of temperature $\theta$ and pressure $P$ go as $\langle \theta\rangle \sim r^{2/3}$ and $\langle P\rangle \sim r^{4/3}$ as a function of distance $r$ \citep{paniveni2010activity}. 
 
For the isosurface of a quantity $X$, the fractal dimension is given by $D_{X} = D_{\rm Euclid} - \frac{1}{2} \times (\textrm{Exponent~of~} \langle X\rangle)$, where the angles indicate variance. For an isothermal supergranular boundary, we thus have the fractal dimension $D_{\theta} = 2 - \frac{1}{3} \approx 1.67$, whereas in the isobaric case, we have $D_{P} = 2 - \frac{2}{3} \approx 1.3$. 

In this light, the data in Table \ref{tab:inflection} bears out the following features that may underlie the bifractal behavior: in the intermediate phase, the isosurface property shows a marked dependence on scale. Specifically, the boundary of larger cells shows a more isothermal character, whilst that of smaller cells shows a more isobaric character. On the other hand, although the bifractality can be well discerned in the minimum and peak phases, still the former is largely consistent with isothermal cell boundaries, and the latter phase with isobaric boundaries.

If we idealize this dependence of the fractal dimension on the phase $\varphi$ of the Solar cycle as an approximately linear function, then we may propose the following:
\begin{equation}
D = \alpha + \beta\varphi,
\end{equation}
where $\alpha \approx \frac{4}{3}$ and $\beta \approx \frac{2}{3}$. Here the cycle is assumed to evolve from the minimum phase ($\varphi=-\frac{1}{2}$), through the intermediate ($\varphi=0$), on to the maximum phase ($\varphi=+\frac{1}{2}$).  

The fractal dimension dependence on cycle phase has been reported earlier, e.g., \citet{rajani2022solar, 
	chatterjee2017variation}. The bifractal behavior may arise from the interplay of the magnetic pressure $\frac{B^2_{\rm plasma}}{2\mu_0}$ and the plasma kinetic energy density in the supergranules. In smaller supergranules, which have lower lifetimes \citep{srikanth1999chromospheric, sowmya2022supergranular}, the magnetic upflow at the supergranular center does not have time to dissipate, letting the magnetic pressure play a more dominant role in relation the plasma kinetic energy density than in larger cells. Thus the supergranular boundary shows a more isobaric property in the smaller cells.  By contrast, in larger cells, with longer lifetimes, the plasma kinetic energy density dominates, so that isothermal property is more plausible for the boundary.

Applying the Kolmogorov hypothesis for turbulent media to Solar convection, we anticipate that the horizontal velocity $v_{\rm h}$ of supergranular cells  depends on its size $L$ according to
\begin{equation}
v_{\rm h} = \epsilon^{1/3} L^{1/3},
\end{equation}
where $\epsilon$ is the plasma injection rate of the convective cell \cite{sowmya2022supergranular}. Letting $T \equiv L/v_{\rm h}$, the length scale fluctuation $\delta_{L}$, which is a measure of the fractal dimension, and velocity fluctuation $v_{\rm h}$ are related by  $ \equiv T \delta v_{\rm h}$  
\begin{equation}
\delta_{L} = \epsilon^{-1/3} L^{2/3} \delta v_{\rm h}.
\label{eq:deltal}
\end{equation} 
This implies that the fractal dimension should be smaller in more active phase in accordance with the data of Table \ref{tab:inflection}. To give specific numbers, we may let $\epsilon = 2.89 \times 10^{-6}$ km$^2$ s$^{-3}$ and $\delta v_{\rm h} = 74$ m/s \citep{paniveni2004relationship}. Choosing $L \approx 35, 34.5$ and 33.5 Mm for the minimum, intermediate and peak phases, we estimate via Eq. (\ref{eq:deltal}) that $\delta_{L} \approx 5.55, 5.50$ and 5.39 Mm, respectively. This behavior may be attributed to the fact, indicated by a number of works, that magnetic activity anticorrelates with cell size \citep{paniveni2010activity}, which may be attributed to the shrinking effect due to magnetic fields \citep{singh1981dependence, meunier2008supergranules}. 

Eq. (\ref{eq:deltal}) suggests that the bifractal behavior may be because the injection rate and/or the velocity fluctuation shows an increase beyond the point of inflection. Future studies should verify the inflectional behavior, and if verified, shed light on how the cell variables $\epsilon^{-1/3}$ and $\delta v_{\rm h}$  themselves depend on $L$. 


\section{ACKNOWLEDGEMENT} 
We acknowledge the data shared by Jagdev Singh. We are also thankful to Manjunath, Subramanya and Mahesh Koti of PESCE Mandya for help with developing the Matlab program.  
\bibliographystyle{apalike}
\bibliography{./document}
\end{document}